\documentclass[11pt,twoside]{article}
\usepackage{asp2010}
\usepackage{natbib}

\resetcounters

\begin{document}

\title{Slow-Mode Oscillations of Hot Loops Excited at Flaring Footpoints
}
\author{Tongjiang Wang$^{1,2}$, Wei Liu$^{3}$ , Leon Ofman$^{1,2,4}$, and Joseph M. Davila$^2$
\affil{$^1$Physics Dept., Catholic University of America, Washington, DC 20064, USA}
\affil{$^2$NASA Goddard Space Flight Center, Code 671, Greenbelt, MD 20771, USA}
\affil{$^3$Stanford-Lockheed Institute for Space Research, CA 94305, USA}
\affil{$^4$Visiting Associate Professor, Tel Aviv University, Israel}
}

\begin{abstract}
The analysis of a hot loop oscillation event using SOHO/SUMER, GOES/SXI, and RHESSI observations
is presented. Damped Doppler shift oscillations were detected in the Fe\,{\sc{xix}} line by SUMER,
and interpreted as a fundamental standing slow mode. The evolution of soft X-ray emission from
GOES/SXI and hard X-ray sources from RHESSI suggests that the oscillations of a large loop 
are triggered by a small flare, which may be produced by interaction (local reconnection) 
of this large loop with a small loop at its footpoint. This study provides clear evidence 
supporting our early conjecture that the slow-mode standing waves in hot coronal loops are excited 
by impulsive heating (small or microflares) at the loop's footpoint.
\end{abstract}

\section{Introduction}
\begin{figure}[!ht]
\plotone{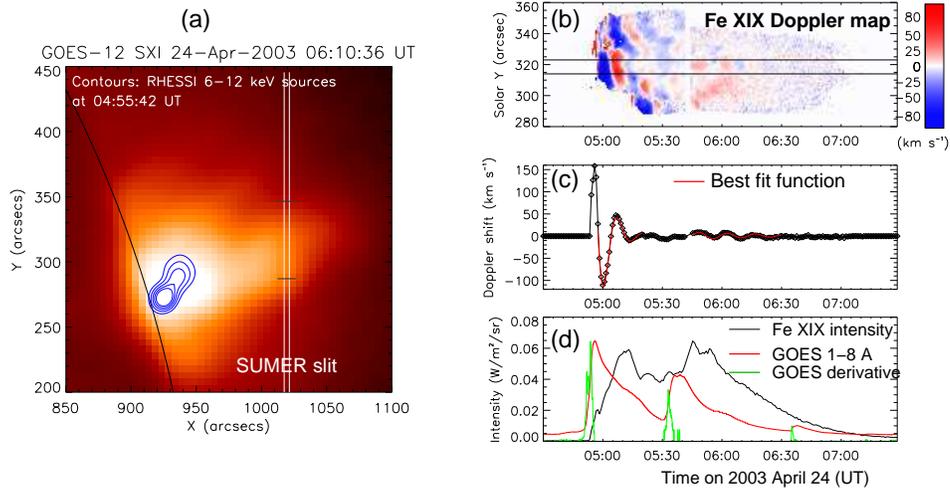}
\caption{Observations of a hot loop oscillation event on 2003 April 24. (a) GOES/SXI image overlaid
with RHESSI 6-12 keV X-ray sources (in contours at 0.3,0.4,0.5,0.6,0.7 of the peak flux). (b) Time series
of Doppler shifts in Fe\,{\sc{xix}} from SUMER. (c) Time profile of Doppler shift oscillations averaged at
the marked position in (b), with the best fits by a damped sine function (in red). (d) Time profile of
the Fe\,{\sc{xix}} intensity (in black) at the marked position in (b), and light curves of
GOES 1-8 \AA\ X-ray flux (in red) and its derivative (in green) in arbitary units. \label{fig}}
\end{figure}

A large number of strongly damped oscillations in hot ($>$6 MK) coronal loops
have been observed by SOHO/SUMER in the past decade in Doppler shifts
of flare lines (Fe XIX and Fe XXI) \citep{wan02, wan03a}. These oscillations with
periods on the order of 10-30 min have been interpreted
as fundamental standing slow modes \citep{ofm02, wan03b}. They often manifest
features such as recurrence and association with a flow (100-300 km/s)
pulse preceding the oscillation, which suggests that they are most likely
driven by microflares at the footpoints \citep{wan05}. These observations
provide important constraints that can be used for improving theoretical models of 
magnetosonic wave excitation and for coronal seismology \citep[][for recent review]{wan07,wan11}.
Here we study an event on 2003 April 24 in AR 10339 (Fig.~\ref{fig}a) to explore its trigger 
using GOES/SXI and RHESSI data.

\section{Observations}
By examining 30 oscillation events observed by SUMER at solar limb during 2002-2003, 
hard X-ray sources are found in the initial phases of more than 10 events.
However, due to the loss of {\it Yohkoh}, no SXT images can be used to identify 
the spatial relationship between the flare sources and the oscillating loop. Instead,
in this work we analyze data from the Solar X-Ray Imager (SXI) onboard {\it GOES-12}.
The coalignment between GOES/SXI and RHESSI images is
based on SOHO/EIT 195 \AA\ images.

\section{Results and Discussion}
Two oscillations are observed (Figs.~\ref{fig}b-d). One follows a GOES C6.7 flare at 04:56 UT 
showing strong damping, and the second follows a C4.5 flare at 05:40 UT showing weak damping.
We measured the first oscillation with the period of 13.5 min, decay time of 6.2 min,
and velocity amplitude of 189 km~s$^{-1}$, and the second one with the period of 14.2 min,
decay time of 44.3 min, and amplitude of 8.2 km~s$^{-1}$. In the late phase, since more mass
has been loaded in the loop by the chromospheric evaporation, it has the high density,
and so oscillates with the smaller amplitude and weak damping.
The RHESSI hard X-ray sources appear as a small loop structure, located at one footpoint
of a large soft X-ray loop, suggesting that the flares and oscillations are triggered by
their interaction (via local magnetic reconnection). 

\acknowledgements 
TW and LO are supported by NASA grants NNX08AE44G, NNX10AN10G, and NNX09AG10G, 
and WL is supported by AIA contract NNG04EA00C.

\bibliography{wangt}

\end{document}